\documentstyle[pre,aps,epsfig]{revtex}
\begin{document}
\draft
%\twocolumn
\twocolumn[\hsize\textwidth\columnwidth\hsize\csname@twocolumnfalse\endcsname
\title{Breakdown of Dynamical Scale Invariance in the
Coarsening of Fractal Clusters}
\author{Massimo Conti}
\address{Dipartimento di Matematica e Fisica, Universit\'{a} di Camerino,
and Istituto Nazionale di Fisica della Materia, 62032,
Camerino, Italy}
\author{Baruch Meerson}
\address{The Racah Institute  of  Physics, Hebrew
University   of  Jerusalem,
Jerusalem 91904, Israel}
\author{Pavel V. Sasorov}
\address{Institute of Theoretical and Experimental
Physics, Moscow 117259, Russia}
\maketitle
\begin{abstract}
{We extend a previous
analysis [PRL {\bf 80}, 4693 (1998)] of breakdown of dynamical scale
invariance
in the coarsening of two-dimensional DLAs (diffusion-limited aggregates)
as described by the Cahn-Hilliard equation.
Existence of a second
dynamical length scale,
predicted earlier, is established. Having measured the ``solute mass"
outside the cluster
versus time,
we obtain a third dynamical exponent. 
An auxiliary problem
of the dynamics of a slender bar (that acquires a
dumbbell shape)
is considered.
A simple
scenario of coarsening of fractal clusters
with branching structure is suggested that employs
the dumbbell dynamics results. This scenario
involves two dynamical
length scales: the characteristic width and length of the cluster
branches. The predicted dynamical exponents
depend on the (presumably invariant) fractal dimension
of the cluster skeleton. In addition, a robust theoretical
estimate for the 
third dynamical exponent is obtained.
Exponents found numerically are in reasonable
agreement with these predictions.}
\end{abstract}
\pacs{PACS numbers: 61.43.Hv, 64.60.Ak, 05.70.Fh}
\vskip1pc]
\narrowtext
%\pagebreak
\section{Introduction}
Nonlinear dissipative
systems, driven out of equilibrium, decay to an
equilibrium state after
the driving agent is switched off or depleted.
As long as a freely decaying nonlinear
system is far from equilibrium, the
relaxation dynamics
are non-trivial and
it is natural, in simple cases, to look for
dynamical scaling and universality. A wide
class of nonlinear relaxation
problems appears in the
context of
phase ordering dynamics
\cite{Gunton,Bray}. In the present work we explore a new
aspect of phase ordering in systems with a conserved order
parameter. This aspect
appears when the minority phase
has long-range correlations and represents (at least at early times)
a fractal cluster (FC). Although such an initial condition
does not result
from a quench from high to low temperature (a standard setting
of phase-ordering dynamics), it is by no means
artificial. There are
many two-phase systems that
exhibit morphological
instabilities and ramified growth at an early stage of their dynamics,
and coarsening at a later stage.
A canonical
example is provided by diffusion controlled
systems, such as an overcooled liquid or super-saturated solution. The stage
of morphological instability and its implications in this system
have been under
scrutiny
\cite{Langer,Kessler,Brener,Mineev,BMT}. If some noise is
present, a DLA-like
FC develops at
this stage \cite{BMT}. The
subsequent surface-tension-driven
coarsening of this FC
is unavoidable if the system is isolated so that the total
amount of available mass or heat is finite. Until recently, this later stage
had received only a limited
attention. Irisawa {\it et al.}  \cite{Irisawa95} carried out
Monte-Carlo simulations
of diffusion-controlled
coarsening of a two-dimensional DLA
cluster and found
a power law with a non-trivial
exponent for the cluster perimeter
as a function of time. More recently the first results
of investigation of the coarsening of two-dimensional DLA clusters
as described by the
Cahn-Hilliard (CH) equation were reported \cite{CMS}.
These results are briefly reviewed in the following.

A crucial issue in the theory of phase ordering processes is the presence
(or absence) of dynamical
scale invariance  \cite{Gunton,Bray}. Dynamical scale
invariance implies that
there is a single dynamical
length scale $\lambda (t)$ such that the coarsening system
looks (statistically) invariant in time when lengths are scaled by
$\lambda (t)$. It was found
Ref. \onlinecite{CMS} that
dynamical scale invariance breaks down during the coarsening
of DLA clusters as described by the CH-equation. On the other hand,
the coarsening dynamics apparently
exhibit {\it scaling}, by which we mean
power laws in time. These power laws (with non-trivial exponents)
were found for the cluster perimeter \cite{Irisawa95,CMS}, and
for the dynamical length scale
that shows up in the
Porod-law part of the equal-time correlation function \cite{CMS}. We will
call this dynamical length scale ``the first correlation
length". The absolute values of the
dynamical exponents for the cluster perimeter and
for the
first correlation length
are close to each other. The
gyration radius of the cluster was found to be constant
(within possible logarithmic corrections). The
mass dimension of the cluster
was also found to
be constant (on a shrinking interval of distances).
The last two findings indicate that the
mass transport
is essentially local at this stage of coarsening.
We conjectured in Ref. \onlinecite{CMS}
that an additional dynamical length scale (with an
exponent larger than that of
the first correlation length scale) must show up in the coarsening
morphology. Furthermore, we speculated that the two different
dynamical length
scales are the average
{\it width} and {\it length}
of the cluster branches.

Fractal coarsening occurs in many
physical systems. Two-dimensional
fractal fingering, observed in a Hele-Shaw cell with
radial geometry (for a recent review see Ref. \onlinecite{fingering}),
exhibits coarsening at a late
stage of the experiment, when
forcing of the more viscous fluid by the
less viscous fluid stops.
Fractal coarsening has been under scrutiny in the context of
sintering, in particular of
silica aerogels \cite{Sempere,Hinic,Jullien}.
Additional examples are provided
by thermal relaxation of initially fractal
grain boundaries \cite{Streitenberger} and by smoothing
of fractal polymer structure in the process of polymer collapse
\cite{Crooks}. It is interesting, to what extent
fractal coarsening is universal.

In this
paper we significantly extend the analysis of Ref. \onlinecite{CMS} of
the
coarsening of DLA fractals as described by the CH equation.
In Section 2 we will report numerical evidence for 
the existence of an additional dynamical
length scale and find the 
corresponding (second) dynamical exponent.
We also introduce in Section 2 an additional measure of the
coarsening dynamics:
the total ``solute mass" content outside the cluster, and find the
corresponding (third) dynamical exponent.
In the rest of the paper we will try to develop some
theoretical understanding of
our numerical results. To this end,
we consider, in Section 3,
the coarsening dynamics
of a single slender bar. The results
of this analysis are employed
in  Section 4, where
a simple  scenario of
coarsening of a FC having branching structure
is suggested. In this scenario
two different dynamical
length scales are present: the characteristic
width and length of the cluster branches. The corresponding
dynamical exponents are calculated; they are found to
depend on the
(presumably invariant)
fractal dimension $D$ of the cluster ``skeleton". A 
robust theoretical
estimate is also obtained for the third 
(``solute mass") dynamical exponent. These
predictions are in agreement with
our numerical
results. A change of sign of the third dynamical exponent is
predicted at a critical fractal dimension $D_{crit}=4/3$. Section 5 includes
a summary and discussion.

\section{Coarsening of DLA clusters: numerical results}
We start with
a brief
description of our simulations and diagnostics. We solved the (dimensionless)
CH equation
\begin{equation}
\frac{\partial u}{\partial t} +
\frac{1}{2}\nabla^2 \left(\nabla^2 u + u - u^3 \right) = 0
\label{f5}
\end{equation}
numerically by discretizing it on the domain $\Omega$:
$0 \le x \le 512\,, 0 \le y \le 512$ with periodic boundary conditions.
An explicit Euler integration scheme was used to advance the solution
in time, and second order central differences to discretize the Laplace
operator. With a mesh size $\Delta x = \Delta y = 1$ no preferred directions
emerged in the computational grid, due to the truncation errors;
a time step $\Delta t = 0.05$ was required for numerical stability. 

\begin{figure}[h]
\vspace{0.5cm}
\hspace{-1.6cm}
\rightline{ \epsfxsize = 6.5cm \epsffile{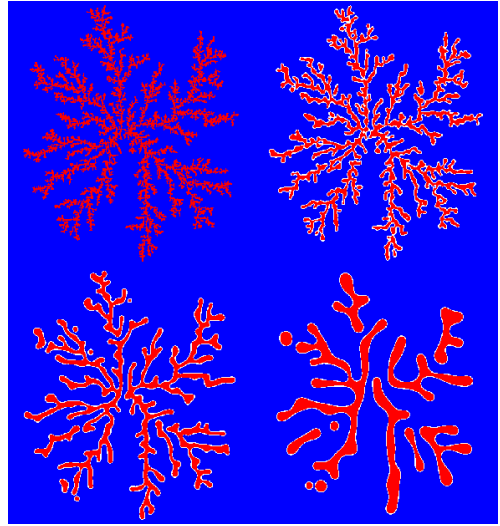}}
\vspace{0.5cm}
\caption{
Coarsening of a DLA fractal cluster
as described by the CH-equation. The upper
row corresponds to $t=0$ (left) and $34.7$ (right), the lower row to
$t=329.3$ (left) and $4,900$ (right).
\label{fig0}}
\end{figure}

DLA clusters \cite{Witten} (like the one shown in
Fig.~\ref{fig0}, upper left), with radius of order 250, were
prepared by a standard random-walk algorithm on a two-dimensional square
grid and
served as the initial conditions
for the minority
phase $u=1$. To prevent breakup of the clusters at an early stage of the
coarsening process, we followed
the technique of Irisawa {\it et al.} \cite{Irisawa95} and
reinforced
the aggregates by an addition of peripheral
sites. Comparing the correlation functions
before and after the reinforcement,
we verified that the reinforcement did not spoil the fractal
properties of the cluster.

It is convenient to introduce
the density field
$$\rho ({\bf r}, t) = \frac{u ({\bf r}, t)+1}{2}\,,$$
which varies between $0$ and $1$.
We identified
the cluster as the locus where $u ({\bf r}, t) \ge 0$, or
$\rho ({\bf r}, t) \ge 1/2$.
The coarsening process was followed up to a time $t=5,000$.
Typical snapshots of the coarsening process are shown in Fig.~\ref{fig0}.
One can see
that smaller features of the FC
are ``consumed" by larger features, while the large-scale
structure
of the cluster is not affected.

To characterize the coarsening process,
several
quantities were sampled and averaged over 
10 initial
configurations. In this paper we will analyze the
following quantities:
\begin{enumerate}
\item
Gyration radius of the cluster.
\item
Circularly averaged equal-time
pair correlation function, normalized at $r=0$:
\begin{equation}
C(r,t) = \frac{\langle \rho({\bf r}+{\bf r}^{\prime},t)
\rho({\bf r}^{\prime},t)\rangle}{\langle\rho^2 ({\bf r}^{\prime},t)\rangle}\,.
\label{corrfunc}
\end{equation}
\item
An estimate of the cluster
perimeter $P (t)$,
defined as the number of broken bonds between the aggregate sites.
\item
The ``solute mass"
outside the cluster:
\begin{equation}
M_s=\hspace{1em}\int\!\!\!\int\limits_{\hspace{-1em}
\rho ({\bf r},t)<1/2}\,
\rho\,({\bf r},t) \, dx\, dy\ .
\label{Mv}
\end{equation}
\end{enumerate}

Quantities 2 and 3 were computed after binarization of the
phase field data: the value of $\rho=1$ is attributed to 
all sites belonging to the cluster, while $\rho=0$ is 
attributed to the rest of sites. 

As we have already noticed,
the gyration radius of the cluster remains
constant (within possible logarithmic corrections) until the
latest available times, so
we will concentrate
on the rest of the measurements.

\subsection{Equal-time pair correlation function}

An analysis of the $r$-dependence of the equal-time
correlation function at different
times shows that coarsening
operates only at small and intermediate distances. In the
following we
will consider separately
two regions of distances.

\subsubsection{Small distances}

\begin{figure}[h]
\rightline{ \epsfig{file=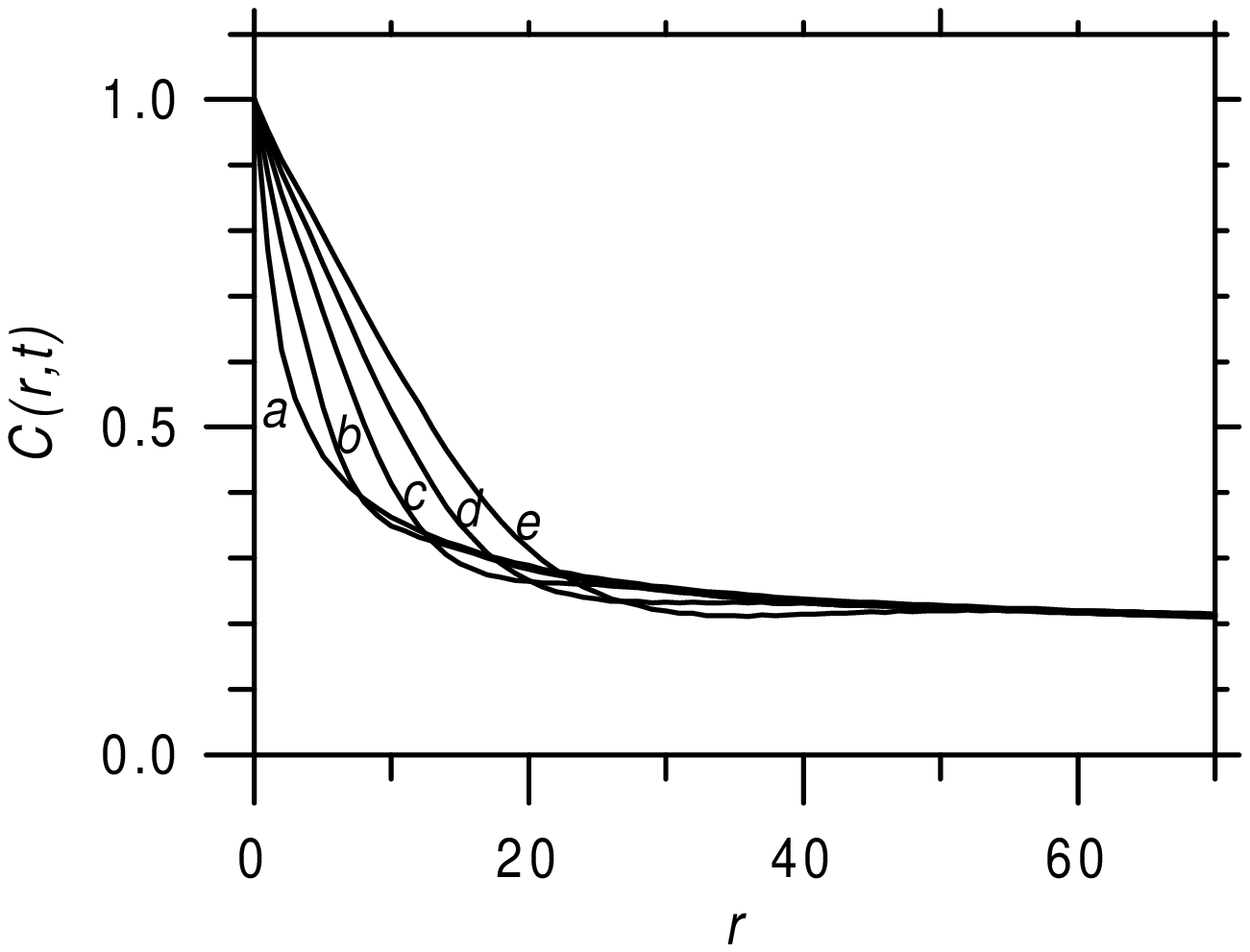, width=3.4in, clip= }}
\vspace{0.2in}
\caption{
Dynamics of the
equal-time pair correlation function $C(r,t)$ at small and intermediate
distances for time moments
$t=0$ (a), $34.7$ (b), $516.5$ (c), $1992$ (d) and $4900$ (e). 
\label{fig2}}
\end{figure}

The 
linear behavior of the correlation function
at small distances (the Porod law), clearly
seen in Fig.~\ref{fig2}, yields the first
correlation length $l_1(t)$
and corresponding dynamical
exponent. In a two-phase
system with a sharp interface, the first correlation
length $l_1$ is the average minimum distance between a randomly chosen
point of the cluster and the interface. We will interpret
$l_1$ as the typical
{\it width} of the cluster's branches (see Fig.~\ref{fig0}).

\begin{figure}[h]
\rightline{\epsfig{file=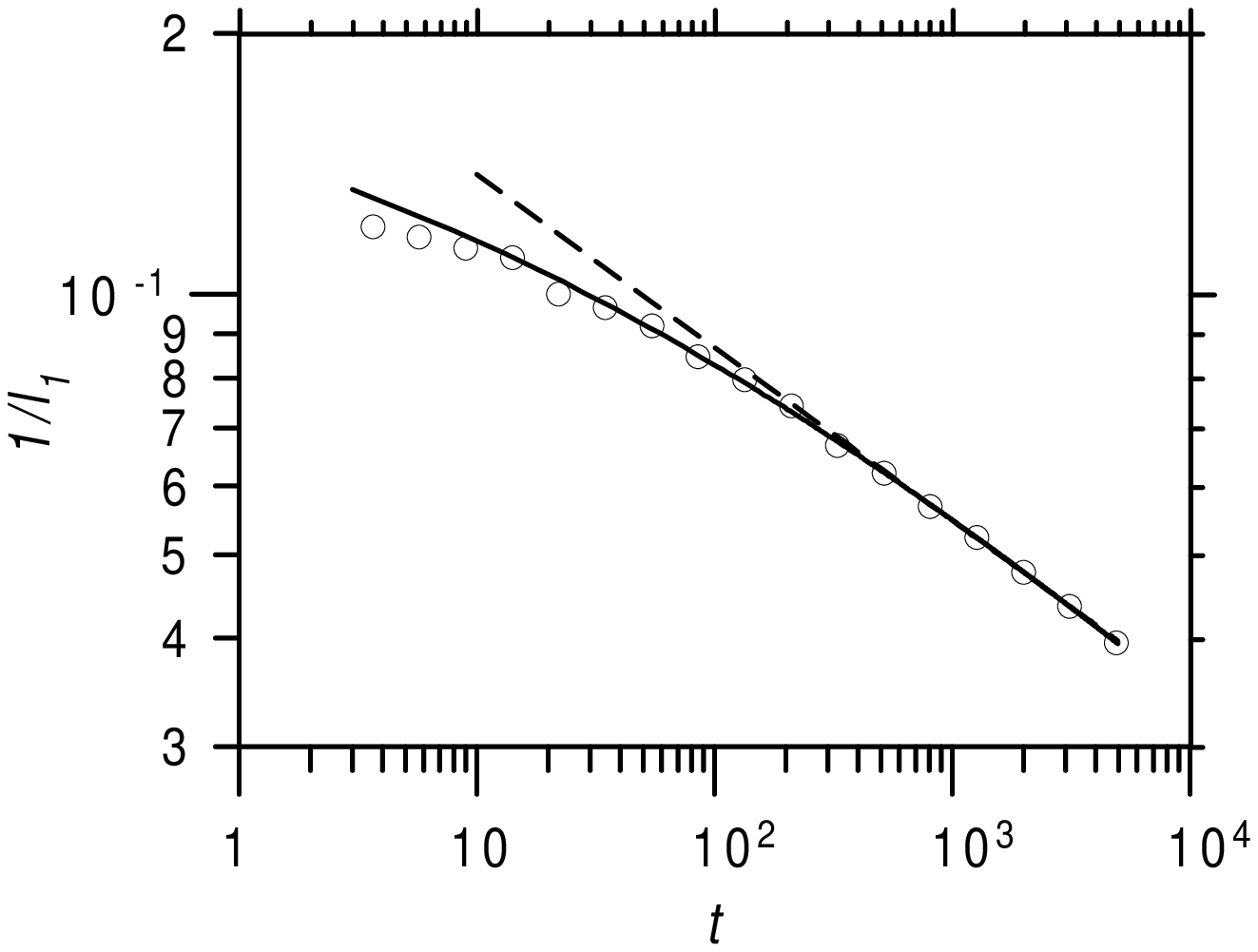, width=3.4in, clip= }}
\vspace{0.2in}
\caption{
The inverse first
correlation length $1/l_1(t)$ versus time (circles)
and its fits (\ref{LC5}) (solid line) and (\ref{LC1}) (dashed line).
\label{fig3}}
\end{figure}

To determine $l_1(t)$, we approximated $C(r,t)$ on the interval
$0.7\le C(r,t)\le 1$ by a linear function $1-r/l_1(t)$.
Fig.~\ref{fig3} shows the inverse correlation length $1/l_1(t)$
versus time, and a corrected-power-law fit
\begin{equation}
\frac{1}{l_1(t)} = \frac{A_{l_1}}{t^\alpha+B_{l_1}}\ ,
\label{LC5}
\end{equation}
of this dependence, with
\begin{equation}
\alpha = 0.26\\ , \,\,\, A_{l_1}=0.44\ \mbox{~~and~~}
B_{l_1}=2.0\ .
\label{LC8}
\end{equation}
This fit was obtained on the interval $30 \le t \le 2000$ that spans from
the time when quasi-equilibrium
sharp interfaces have already formed until the time
when the system size becomes important (see below). Here and in the
following we do not show
the standard deviations of the fitting parameters if they are 
less than or equal to
unity in the last significant digit presented.
The dynamical exponent $\alpha = 0.26$
differs from the Lifshitz-Slyozov value of $1/3$
observed in those cases when phase-ordering processes
exhibit dynamical scale invariance
\cite{Gunton,Bray,LS,RogersDesai,Tomita,Jeppesen,Huse}.
In contrast to such processes,
no {\it a priori} form for the
correction to the power-low fit for $1/l_1(t)$
is available, so we
consider the corrected-power-law
fit (\ref{LC5}) as empirical. It
works well though (see Fig.~\ref{fig3}).

A simpler alternative 
(chosen in Ref. \onlinecite{CMS})
is to use
a pure power-law fit. For the last decade of time,
$490 \le t \le 4900$, it gives
\begin{equation}
l_1(t)= C_{l_1}\, t^{\alpha}\ ,
\label{LC1}
\end{equation}
where
\begin{equation}
\alpha=0.20\mbox{~~~~~~and~~~~~~}C_{l_1}=4.6\ .
\label{LC2}
\end{equation}
The
difference between the values 0.26 and 0.20 is disappointingly large;
it gives a measure of the maximum uncertainty of this exponent,
caused by systematic errors.
Similar uncertainties occur for other dynamical
exponents that we find in this work. 
Our conclusion that
$\alpha$ is smaller than the Lifshitz-Slyozov value
$1/3$ is unaffected
by this uncertainty. In Section 4 we will present
a simple coarsening
scenario that gives a theoretical prediction for
$\alpha$ which agrees, for DLA clusters, with the value 0.26.

\subsubsection{Intermediate distances}

A close inspection of the correlation function $C(r,t)$
at intermediate and large distances show that there {\it are} 
small (within
$5$ to $10$\%) changes there. These changes result from
small systematic variations of the cluster mass
in the process of coarsening
(see below). They
make it difficult to perform accurate measurements at intermediate and large
distances, therefore some normalization at these distances
is necessary.
We
normalized the
correlation function $C(r,t)$ at different moments of time
to its values at $r=120$:
\begin{equation}
\hat{C}(r,t)=\frac{C(r=120, t=0)\,C(r,t)}{C(r=120, t)}\,.
\label{N1}
\end{equation}
The results obtained by using the normalized
function $\hat{C}(r,t)$ are not sensitive to the
exact value of $r$ chosen
for the normalization, as long as it is large enough.

\begin{figure}[h]
\rightline{ \epsfig{file=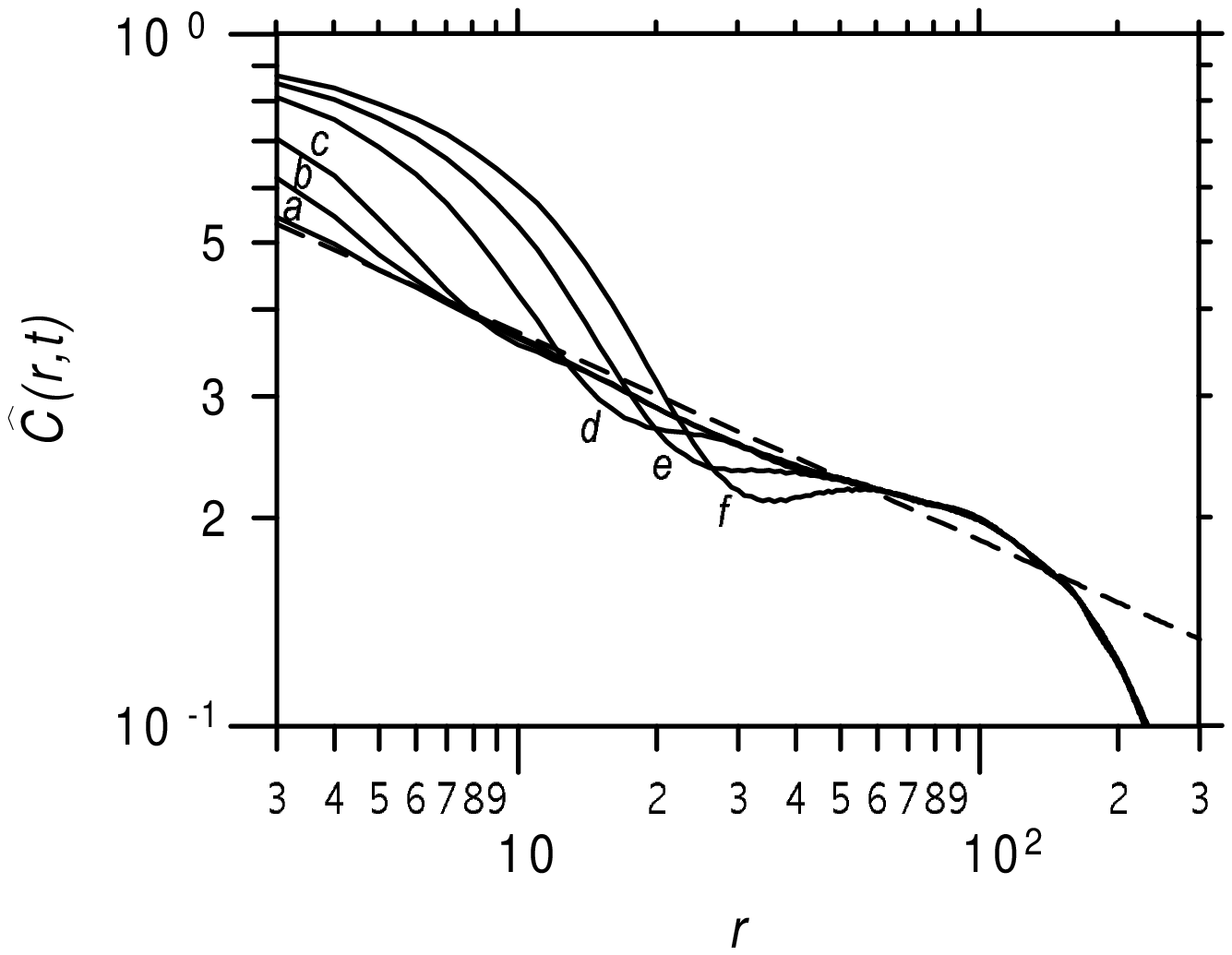, width=3.4in, clip= }}
\vspace{0.2in}
\caption{
Dynamics of the normalized equal-time pair correlation
function $\hat{C}(r,t)$ [Eq. (\ref{N1})]
at time moments $t=0$ (a), $3.65$ (b), $34.7$ (c), $516.5$ (d), $1992$ (e)
and $4900$ (f).
The dashed line shows the power-law fit $0.74\, r^{-0.30}$
on the interval $3\le r \le 150)$.
\label{fig1}}
\end{figure}

Figure ~\ref{fig1} shows $\hat{C}(r,t)$ at different moments of time.
The log-log plot helps to decide
on the range of distances and
times we can work with. It is seen that
$\hat{C} (r,t)$ changes, as a function of time, only at distances smaller
than some
correlation radius $r_c(t)$ that increases with time.
At distances $r>r_c(t)$
$\hat{C} (r,t)$ stays very close to its
initial value $C(r,0)$.  As it is our aim to investigate an intermediate
asymptotic
coarsening regime related to the fractal structure of the
cluster at $t=0$, we should work on
a (shrinking) interval of scales where, at $t=0$, $C(r,0)$ exhibits
a power-law behavior \cite{Vicsek}. A
reasonably accurate power-law
fit can be achieved at $t=0$ on the
interval
$3<r<150$, while beyond $r=L=150$ (the upper cutoff of the DLA cluster)
finite-size effects become large. The fitting function is
$c\, r^{-\delta}$,
where $\delta=0.30$ and $c=0.74$.
Correspondingly, the fractal dimension
of our DLA clusters is
$D=2-\delta=1.70$, a reasonably accurate value in view of
the relatively small size of our system. The
inequality $r_c(t) \ll L$ puts an upper limit
on the coarsening time that we can still work with.
Fig.~\ref{fig1} shows that for
the last available time  of our simulations, $t=4900$, the ratio of
$r_c/L$ is already about $0.3$. Therefore, for more reliable results
we should limit ourselves by $t=2000$, as we have already done in
the fitting of $l_1(t)$.

\begin{figure}[h]
\rightline{\epsfig{file=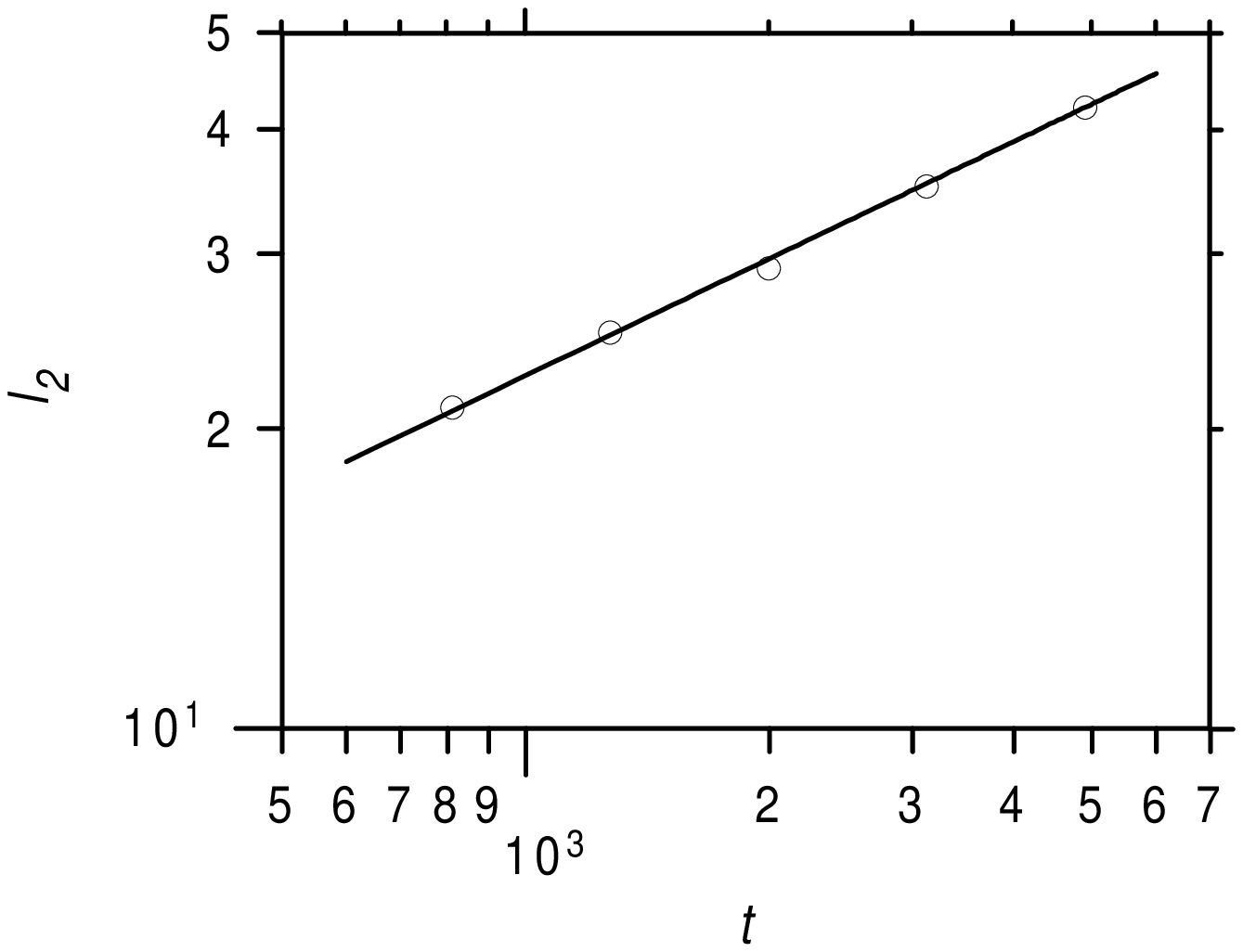, width=3.4in, clip= }}
\vspace{0.2in}
\caption{
The second correlation length $l_2(t)$ versus time (circles),
and
its fit (\ref{LB1}) (solid line).
\label{fig3a}}
\end{figure}

The normalized
correlation
function $\hat{C} (r,t)$ exhibits an additional
dynamical length scale that appears at intermediate distances.
Recall
that, at sufficiently large distances, the
tails of $\hat{C}(r,t)$ at different time moments coincide with
$\hat{C}(r,0)$ (see  Fig.~\ref{fig1}). We define $l_2(t)$ as the
minimum value of
$r_1$ such that for $r>r_1$ $\hat{C}(r,t)$ is less
than $\hat{C}(r,0)$ by no more
than 10\%. Such values of
$l_2(t)$ exist only for $t>800$, and they are shown, on the log-log plot,
in Fig~\ref{fig3a}. As the available time interval in
this case is quite short (less than one decade), we are forced to
use the last available time $t=4900$ and limit
ourselves to a pure power-law fit:
\begin{equation}
l_2 (t)= C_{l_2}\, t^{\beta}\ ,
\label{LB1}
\end{equation}
where
\begin{equation}
\beta=0.39 \mbox{~~~~~and~~~~~}C_{l_2}=1.5 \ .
\label{LB2}
\end{equation}

Though the standard deviation is small, the short time interval
does not guarantee a high precision of
the exponent $0.39$. A big difference between
the exponents $\alpha$ and $\beta$ is, however, beyond doubt.
We will interpret
$l_2(t)$ as the characteristic branch {\it length} which serves as the
time-dependent lower cutoff of the
fractal ``skeleton" of the coarsening cluster.

Equations (\ref{LC2}) and (\ref{LB2}) show an interesting
relationship between $l_1$ and $l_2$:
\begin{equation}
l_1 l_2^2\propto t^{0.98}\ .
\label{LB3}
\end{equation}
The exponent 0.98 is close to unity, and we will return
to this observation in Section 4.

It should be noted that normalization
of $C(r,t)$ at large $r$, which helped us to extract the second dynamical
length scale,
is no more artificial than the widely
used normalization at $r=0$. The only real need for {\it any} normalization
of the equal-time pair correlation function 
is non-constancy of the cluster mass in time.  When 
investigating the dynamics at
small distances,
it is convenient to normalize the correlation function at $r=0$. 
When investigating the dynamics at
large distances, it is convenient to normalize $C(r,t)$ somewhere in
the tail.

\subsection{Cluster Perimeter}

\begin{figure}[h]
\rightline{\epsfig{file=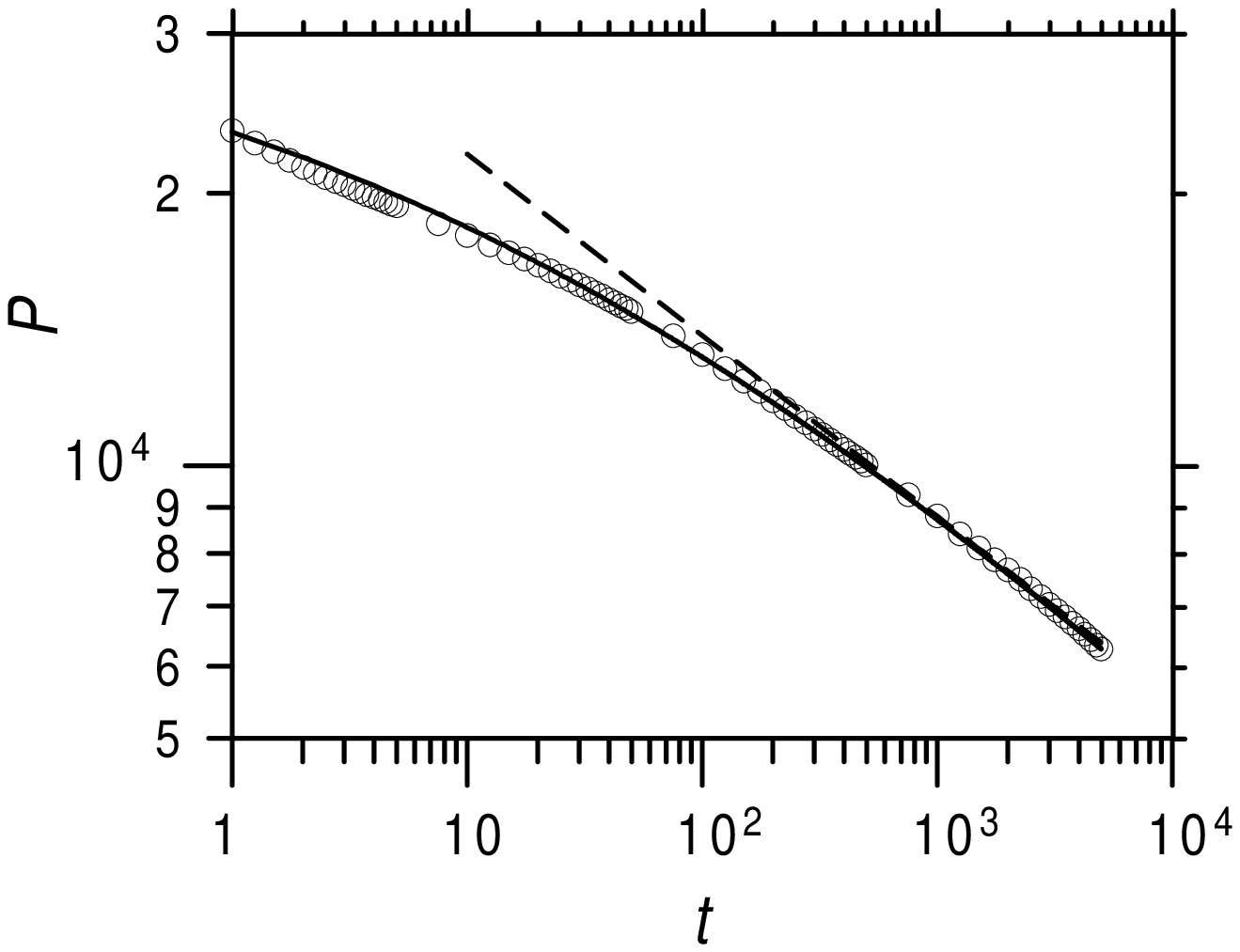, width=3.4in, clip= }}
\vspace{0.2in}
\caption{
Cluster perimeter $P(t)$
versus time (circles) and its fits
(\ref{P01}) (solid line) and (\ref{P03}) (dashed line).
\label{fig5}}
\end{figure}

Figure ~\ref{fig5} shows the dynamics of the cluster perimeter
$P (t)$.  A decrease of the perimeter of a cluster
under condition of (approximate) conservation of the cluster mass is
a clear manifestation of coarsening.
The corrected-power-law fit of $P(t)$
is
\begin{equation}
P (t) \approx \frac{A_{P}}{t^{\alpha_{P}}+B_{P}}\ ,
\label{P01}
\end{equation}
where
\begin{equation}
\alpha_{P}=0. 26 ,\quad A_{P}=7.0\cdot 10^4
\mbox{~~~and~~~} B_{P}=2.0\ .
\label{P02}
\end{equation}

A pure power-law fit for the last decade of time gives \cite{CMS}
\begin{equation}
P (t)=C_{P}\, t^{-\alpha_P}\ ,
\label{P03}
\end{equation}
where
\begin{equation}
\alpha_{P}=0.20\mbox{~~~~and~~~~}C_{P}=3.5\cdot 10^4\ .
\label{P04}
\end{equation}
The absolute values of the exponents $\alpha_P$
and $\alpha$
coincide for the same type of fit. 
The inverse slope of the Porod-law-part of the correlation function
should indeed
scale as the perimeter, because each element of interface which size
is much larger than the distance $r$ contributes independently to
the correlation function \cite{Bray}.

\subsection{``Solute" Mass Outside the Cluster}
\label{VMas}

Let us use, for a moment, the language of physics of liquid
solutions.
As {\it all} of the solute mass
at $t=0$ is concentrated in the DLA cluster,
rapid dissolution of the solute from the cluster edge
occurs first.
The cluster mass will start to decrease with time.
Unless the area fraction of the ``fractal phase" is too small
the dissolution stops,
and the dissolved material
precipitates back on the coarsening cluster \cite{remark}.
The mass of the dissolved material decreases at this stage,
while the cluster
mass slowly increases,
asymptotically approaching a constant value.
The late stage of this regime should be qualitatively similar  to
Ostwald ripening, where the dynamics of the 
dissolved material is responsible
for a correction to the Lifshitz-Slyozov scaling behavior \cite{LS}.

%\begin{figure}[h]
%\rightline{ \epsfig{file=FIG7.eps, width=3.4in, clip= }}
%\vspace{0.2in}
%\caption{
%Mass of  ``solute" outside the cluster
%$M_s$ versus time (circles)  and its fits
%(\ref{VM3}) (solid line) and (\ref{VM1}) (dashed line).
%\label{fig6}}
%\end{figure}

\begin{figure}[h]
\rightline{ \epsfig{file=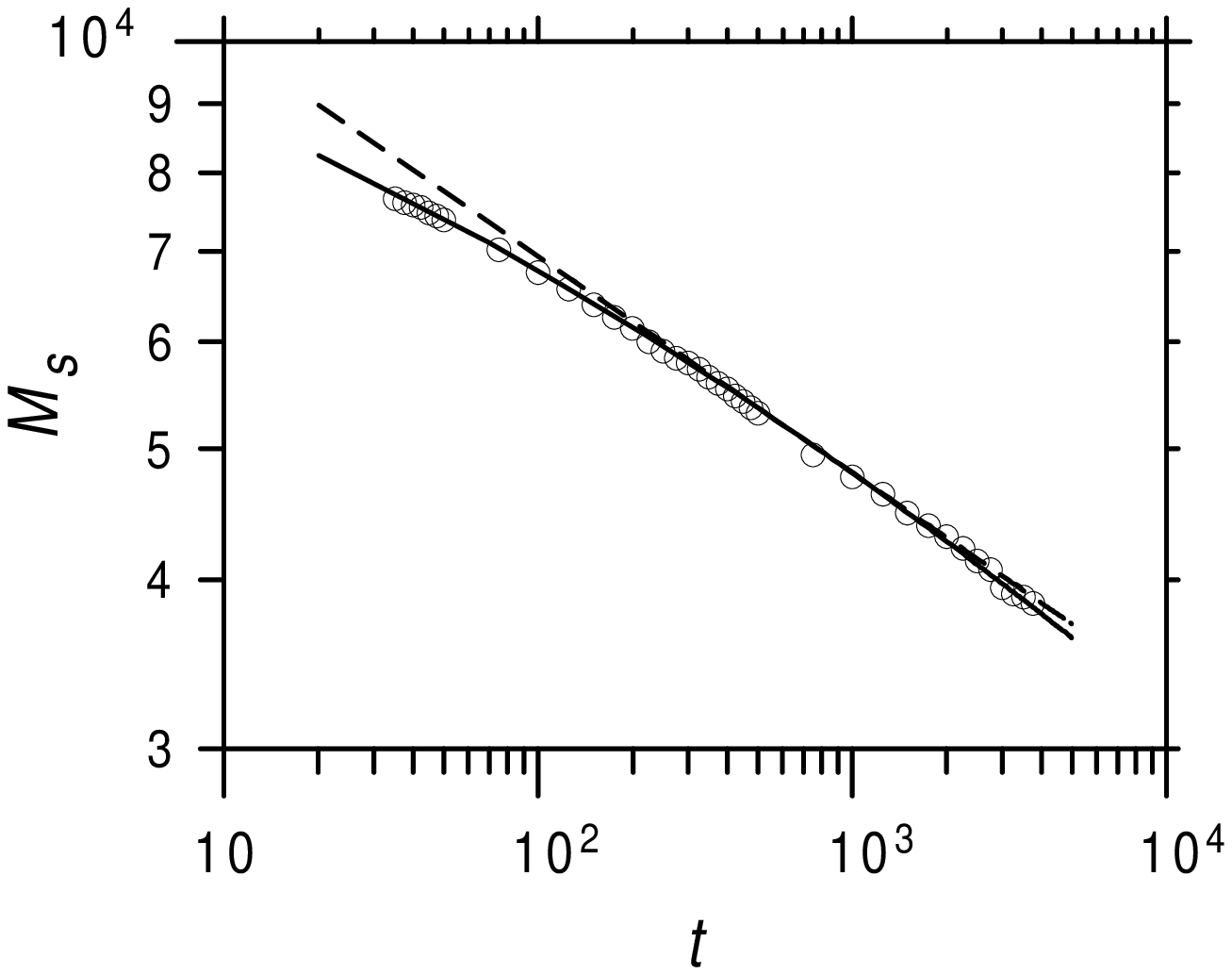, width=3.4in, clip= }}
\vspace{0.2in}
\caption{
Mass of  ``solute" outside the cluster
$M_s$ versus time (circles)  and its fits
(\ref{VM3}) (solid line) and (\ref{VM1}) (dashed line).
\label{fig6}}
\end{figure}

These arguments give a qualitative explanation to our numerical
results on the solute mass outside the cluster versus time, $M_s(t)$.
 Fig.~\ref{fig6}
shows a log-log plot of $M_s(t)$
at sufficiently late times
(the rapid ``dissolution" observed at earlier times
is not shown). The same figure
shows two fitting
functions to $M_s (t)$. A corrected-power-law fit is
\begin{equation}
M_s(t)=\frac{A_{M_s}}{t^{\gamma}+B_{M_s}}\ ,
\label{VM3}
\end{equation}
where
\begin{equation}
\gamma=0.24\,, \quad A_{M_s}=3.64\cdot 10^4\mbox{~~~~and~~~~}
B_{M_s}=2.4 \ .
\label{VM4}
\end{equation}

A pure power-law fit for $500<t<4900$ is
\begin{equation}
M_s(t)= C_{M_s}\, t^{-\gamma}\ ,
\label{VM1}
\end{equation}
where
\begin{equation}
\gamma=0.16 \mbox{~~~~~and~~~~~}
C_{M_s}=(1.45\,\pm 0.02)\cdot 10^4\,.
\label{VM2}
\end{equation}

One can expect that the fractal coarsening dynamics
look simplest (that is, corrections to dynamical scaling are small)
when the cluster mass is already
almost constant.
This condition is only weakly satisfied
in our simulations. For example, even at $t=2000$ the solute
mass outside the cluster
reaches about
10\% of the cluster mass. This may be the reason for the relatively large
uncertainties of the numerical values of the dynamical exponents that we have
found. Another possible reason is the finite size effects related to a moderate
fractal range of the DLA-clusters that we used in simulations.

\section{Slender Bar Dynamics}

In the rest of the paper we will try to get some
quantitative
understanding of our numerical results. As no satisfactory theory for
coarsening of FCs
is available, one can try to formulate a
simplified scenario
and compare its predictions with simulations. The simplest
possible scenario assumes dynamical
scale invariance, that is the presence of a single relevant
dynamical length scale
\cite{Toyoki,Sempere}.
We have seen that, in the case
of CH-dynamics, this scenario disagrees with simulations. 
Breakdown of dynamical scale invariance is caused by the effective
locality of mass transfer which manifests itself in the (approximate)
conservation
of {\it both} the mass, {\it and}
gyration radius of the cluster in the process
of coarsening. 

Looking for a scenario with broken dynamical scale invariance one should,
first of all,  identify the nature of the second dynamical
length scale. Snapshots
of the coarsening process (Fig.~\ref{fig0}) indicate that
it
might be 
the average {\it length} of the cluster branches \cite{CMS}.
Having made this assumption,
we should verify that the typical
width and length of the cluster branches indeed
show different dynamical scalings,
the length growing faster than the width.

An important element of the
scenario that
we want to explore is the shrinking dynamics
of a single slender bar of phase
$u=1$ (``solid"), evolving under the CH-equation (\ref{f5})
in the ``liquid" phase $u=-1$. Special
simulations 
show (see below) that,
in the process of shrinking along its main axis,
the bar acquires the shape of
a dumbbell, and the ``balls" at the ends of the dumbbell
expand with time. In this Section we will derive scaling relations for
the time-dependent parameters of the shrinking dumbbell-shaped bar, and
then compare them with numerical simulations.

\subsection{Slender bar dynamics: theoretical estimates}

Our approach to the slender bar dynamics
employs a modified version of the
rigorous asymptotic sharp-interface theory developed for
the CH-equation \cite{Bray,Pego}. This theory (which can be called
``Laplacian coarsening") requires
all characteristic length scales in the problem
to be much larger than the domain wall width (which
is of order unity), but much 
smaller than the characteristic
diffusion length $l_d \sim t^{1/2}$. Under these assumptions,
\begin{itemize}
\item{the local normal velocity of
the moving interface of the bar is equal to
the difference between the normal components
of the gradient of the density field $\rho ({\bf r}, t)$
outside and inside the
interface,}
\item{the density field is represented by
two harmonic functions: one outside, the other inside the bar,}
\item{there is a Gibbs-Thomson matching condition at
the moving interface and no-flux condition at the external boundary.}
\end{itemize}
This  formulation 
enforces exact conservation of the bar area in the process of 
coarsening. On the other hand, the condition of an 
infinite diffusion length 
is too restrictive. It is not satisfied 
in any of the state-of-the-art
numerical simulations of phase ordering as described by 
the Cahn-Hilliard equation. Fortunately, it is sufficient to
require in practice 
that the diffusion length be larger than the characteristic
coarsening length(s) of the problem, and this is what we shall do.

We assume that
the bar length is much larger than its width and use the aspect ratio
of the bar as a large parameter. Limiting ourselves to
order-of-magnitude estimates, we will assume that
the density field inside the bar has already settled down, so
that we can
treat the bar interior simply as a region where $\rho({\bf r}, t)=1$.

What is the physical picture of the bar dynamics?
Because of the large aspect ratio of the bar,
the density gradient will be
largest near the dumbbell ends and small elsewhere [see Eq. (\ref{B1}) below].
The solute, released from the bar ends, will therefore be
transported to the liquid, and
the bar will be shrinking
along its main axis. Let us assume that
the total mass of the solute outside
the bar is relatively small (see below). Then
most of the dissolved solute will be reabsorbed
not far from the bar ends. This effectively local dynamics
is the reason
that the bar acquires the shape of a dumbbell.
The
``balls" at the ends
of the
shrinking dumbbell will effectively travel along the main axis of the
bar like snowballs, accumulating material along their
motion and
growing in size.

\begin{figure}
\rightline{ \epsfig{file=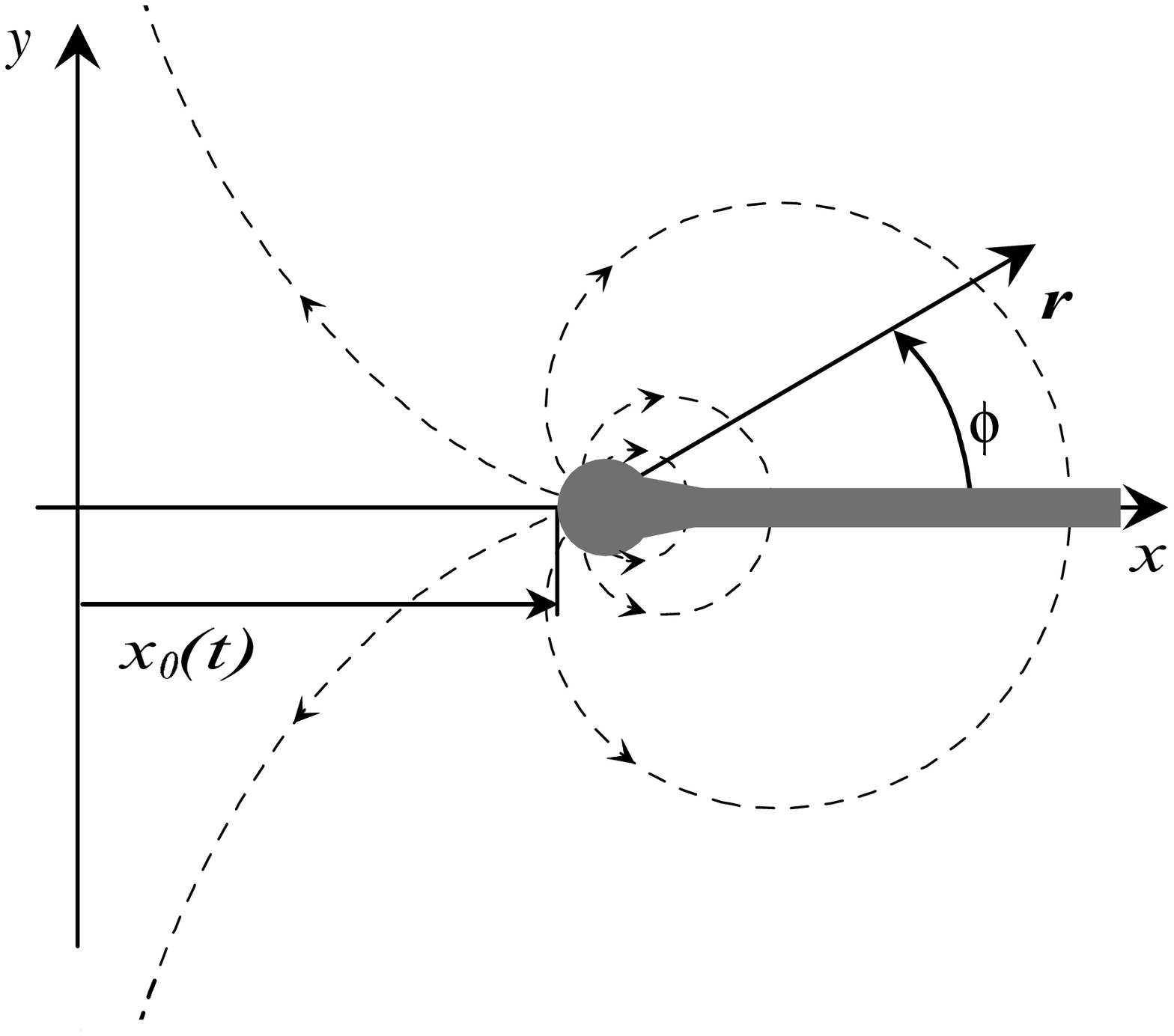, width=3.4in, clip= }}
\vspace{0.2in}
\caption{Setting for the dumbbell dynamics. Dashed lines are
the lines of flow calculated from Eq. (\ref{B1}). }
\label{figbar}
\end{figure}

Going over to a quantitative analysis, we consider
a half-infinite bar
and denote the
width of its planar part by $\Delta$. We
assume that, despite
the fact that the ball at the end of the dumbbell
may have a complicated shape,
it can still be characterized by a single time-dependent
length scale $R(t)$. (Recall that what we are after is
order-of-magnitude estimates, rather than a complete solution.)
Let the bar be placed
along the $x$-axis as shown in Fig.~\ref{figbar}. In this Figure
$x_0 (t)$ is the time-dependent position of the bar edge (that is,
the length reduction of the bar in the
process of its shrinking). Introduce the polar
coordinates
$r$ and $\phi$ with the origin at the
bar end.

The density field $\rho$
must be a harmonic function that vanishes at the planar
part of the dumbbell interface (that is, at $\phi \to 0$ and
$\phi \to 2\pi$)
and is of order $1/R(t)$ at the ``ball" interface
(for definiteness, at $\phi= \pm \pi/2$). Far enough from
the bar end,
$r \gg R(t)$, the bar can
be considered as
a thin sheet [we assume that
$\Delta \ll R (t)$]. The
solution of the Laplace equation for a thin sheet is
elementary \cite{Jackson}, and we can estimate
the density as
\begin{equation}
\rho (r, \phi, t) \sim C \left[R(t)\,r\right]^{-1/2}\,
\sin(\phi/2) ,
\label{B1}
\end{equation}
where $C$ is a constant of order unity.
Employing the boundary condition at the ball interface, we have
extended
the thin-sheet approximation to the limit of its
applicability, but this
can only affect the value of the constant $C$. Dashed lines 
in Fig. \ref{figbar} show the streamlines of the mass flow
obtained by taking the gradient of the density field (\ref{B1}).

Calculating the total mass of the ``solute" with
Eq. (\ref{B1}), one can see that it diverges at large $r$. This divergence 
arises because
one replaces
a diffusion equation by the simpler Laplace equation in the
asymptotic theory of the CH-equation \cite{Bray,Pego}. The divergence can
be
cured by introducing an upper cutoff $R_{max}$
in Eq.~(\ref{B1}). The cutoff is the
smallest of the two lengths: the diffusion length
$l_d \sim t^{1/2}$ and the system size $L_0$.

There are two important
consequences of Eq.~(\ref{B1}).
First, the density gradient $|\nabla \rho|$ is indeed
largest near the bar end. As the result, the ball
emits material from its edge
and retreats. 
Second, the total
mass flux to the planar part of the bar is finite, as it is
proportional to $r^{-3/2}$ and therefore
converges at
large $r$ along the bar edge. The second property
implies
that most of the solute, emitted by the bar end, is reabsorbed
by the less curved part of the ``ball".  Again,
this requires that
the solute mass outside of the bar is small, and
we will return to this condition later.

Now we proceed to obtain simple scaling relations for the
parameters of the shrinking half-infinite
bar. Employing the (approximate) constancy of the bar mass, we
have
\begin{equation}
x_0(t)\Delta\sim [R(t)]^2.
\label{B1a}
\end{equation}
The mass flux out of the ball
can be estimated as
\begin{equation}
\dot{m}\sim |\nabla \rho|_{r\sim R}\, R\sim[R(t)]^{-1}\ .
\label{B2}
\end{equation}
As a result of the outflow and reabsorption of this material
the ball
travels along the
$x$-axis. The characteristic time $\tau_{R}$ it takes the ball to pass
the distance comparable to its size
is $\tau_R\sim m/\dot{m}\sim [R(t)]^3$, so the ball speed is
\begin{equation}
\dot{x}_0(t)\sim\frac{R(t)}{\tau_R}\sim[R(t)]^{-2}\ .
\label{B3}
\end{equation}
Equations (\ref{B1a}) and (\ref{B3}) yield
\begin{equation}
\dot{x}_0(t)\,  x_0(t) \, \Delta  \sim 1\ ,
\label{B4}
\end{equation}
which follows
\begin{equation}
x_0(t)\sim \Delta^{-1/2}\, t^{1/2}\ .
\label{B5}
\end{equation}
Then Eq. (\ref{B1a}) yields
\begin{equation}
R(t) \sim \Delta^{1/4}\, t^{1/4}\ .
\label{B5a}
\end{equation}

Therefore, shrinking of a slender bar in the CH-equation
exhibits dynamical scaling with exponents $1/2$ (for the bar length)
and $1/4$ (for the ball size). In the following
we will verify Eq. (\ref{B5}) numerically and then employ it in our
fractal coarsening scenario.

Let us return to the condition of the relative smallness of the
solute mass outside the bar. Using Eq. (\ref{B1})
with an upper cutoff, we can estimate the solute mass outside the bar
as $[R(t)]^{-1/2}\, R_{max}^{3/2}$. This should
be much less that the ball mass:
\begin{equation}
[R(t)]^{-1/2}\, R_{max}^{3/2} \ll [R (t)]^2\ .
\label{B1b}
\end{equation}
At not too large times, $t < L_0^2$, it is the diffusion length
$l_d \sim t^{1/2}$ that should be taken for the upper cutoff $R_{max}$.
In this regime the
condition (\ref{B1b}) can be rewritten  as
\begin{equation}
\Delta \gg t^{1/5}\,.
\label{B1c}
\end{equation}
At later times, $t > L_0^2$, we should put $R_{max}\sim L_0$, and the condition
(\ref{B1b}) becomes
\begin{equation}
\Delta \gg L_0^{12/5}\, t^{-1}\,.
\label{B1d}
\end{equation}
The criteria (\ref{B1c}) and (\ref{B1d}) coincide (by order of magnitude)
in the intermediate regime $t \sim L_0^2$ and yield
\begin{equation}
\Delta \gg L_0^{2/5}\ .
\label{B1e}
\end{equation}
Eq. (\ref{B1e}) is the most stringent criterion,
so we should require that it holds. On the
other hand, we should assume that an initially rectangular
bar has already developed its dumbbell shape. This requires
$R(t) \gg \Delta$. Using Eq. (\ref{B5a}),
we can rewrite it as $\Delta \ll t^{1/3} $. Overall,
the dumbbell scalings (\ref{B5}) and (\ref{B5a})
are expected to hold when
\begin{equation}
L_0^{2/5} \ll \Delta \ll t^{1/3}\ .
\label{intermed}
\end{equation}

\subsection{Slender bar dynamics: numerical simulation}

The dynamics of
a slender bar were simulated with the same CH equation
(\ref{f5}). We started with a
rectangular-shaped bar ($u=1$ inside the bar,
$u=-1$ outside). The bar sizes were
$512\times 8$, and the bar was placed in the center
of a $1024\times 256$ rectangular box. The CH-equation (\ref{f5})
was solved with periodic boundary conditions.
Only one quadrant was actually simulated
because of symmetry with respect to the $x$ and $y$ axes.

\begin{figure}[h]
\rightline{ \epsfig{file=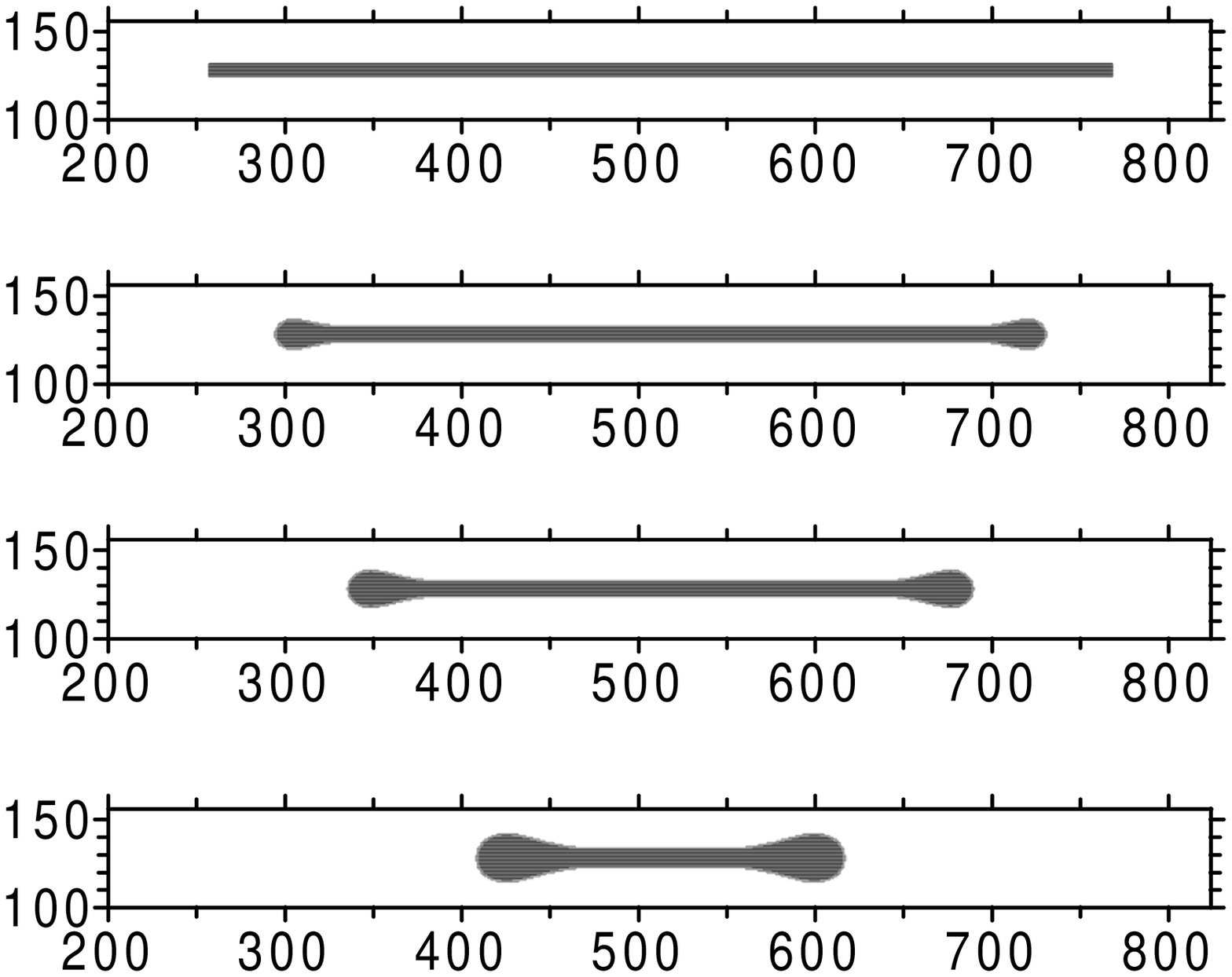, width=3.3in, clip= }}
\vspace{0.2in}
\caption{
Dumbbell formation and dynamics at
$t=0$, 8154, 26810 and 88180 (from top to  bottom).
\label{fig9}}
\end{figure}

Snapshots of the bar evolution
at time moments $t=0$, 8154, 26810 and 88180
are shown in Fig.~\ref{fig9}. They
confirm the ``dumbbell picture" qualitatively. One can see that
most of the
bar keeps its planar shape and constant width.
Balls are formed at the ends of the bar and
travel along the bar's main axis growing in size.

\begin{figure}[h]
\rightline{ \epsfig{file=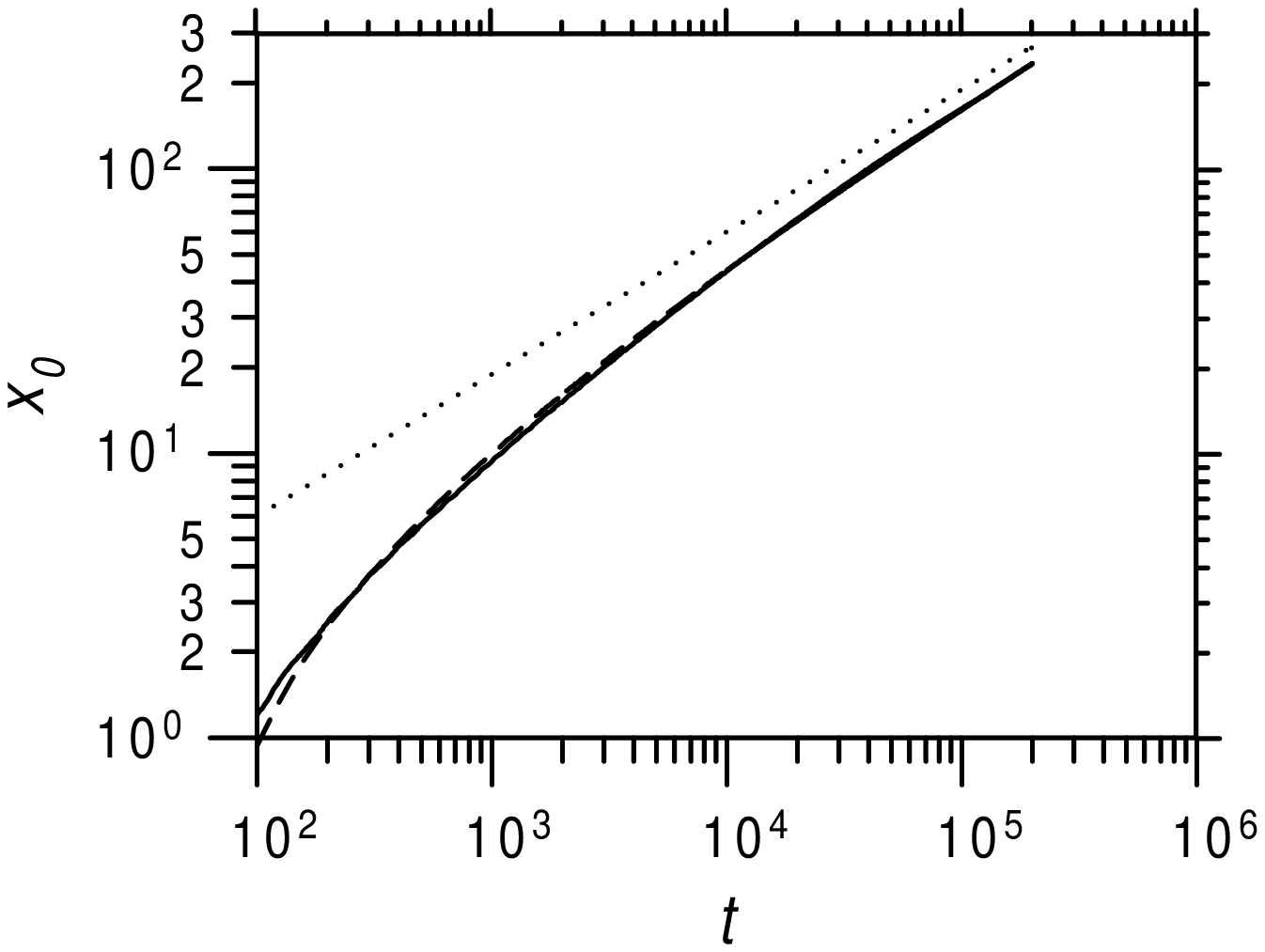, width=3.4in, clip= }}
\vspace{0.2in}
\caption{
The bar length reduction $x_0(t)$
versus time
(solid line); fit (\protect\ref{B21})
(dashed line) and its
leading term (dotted line).
\label{fig10}}
\end{figure}

Fig.~\ref{fig10} shows the numerically obtained time-dependence
of the bar length reduction
$x_0(t)$.
A pure power-law asymptotics of this quantity,
predicted by
Eq. (\ref{B5}), is expected to show up at very late times,
when the typical size of the balls $R(t)$ becomes much larger than
$\Delta$. At intermediate times we can expect a correction of
order $\Delta/R(t)$. For $\Delta=8$ used in our simulations,
this correction
is significant until the latest  available times
$\sim 10^5$. Taking it into account,
we can fit the numerical data for $x_0(t)$
by the function
\begin{equation}
x_0(t) = C_1 \,t^{1/2} + C_2 \,t^{1/4}\ .
\label{B21}
\end{equation}

This fit (with $C_1=0.6$ and $C_2 = -1.6$)
and its leading term
are shown separately
in Fig.~\ref{fig10}, and a good agreement is observed.
Reintroducing the $\Delta$-dependences, predicted by
the theory, we can rewrite Eq.~(\ref{B21})
as
\begin{equation}
x_0(t) = 1.7\, (t/\Delta)^{1/2} - 0.95 \,(\Delta \cdot t)^{1/4}
+ \ldots \ .
\label{B22}
\end{equation}
In
the following Section we will
use only the leading term of
Eq. (\ref{B22}).

The parameters chosen for this simulation made it possible to reach
the dumbbell
scaling regime. Indeed,
the right inequality in Eq. (\ref{intermed}) is satisfied for any of the
time moments shown in Fig. \ref{fig9}, except $t=0$. Taking 
the system length $L_0=124$
(the distance between the planar part of the bar and the boundary), we
see that the left inequality in Eq. (\ref{intermed})
requires $\Delta \gg 7$. The results of the bar dynamics simulation for
$\Delta=8$ indicate that a usual (not strong) inequality is
sufficient. We observed
breakup of
the bar into fragments for $\Delta=4$, when this
condition is violated.

The results of our investigation of the slender bar dynamics are interesting
in their own right. One should work with domains
of a simple shape in order to understand the basics of
coarsening dynamics. 
It is
a single spherical droplet of the minority phase
that usually serves as
a ``test object" for coarsening dynamics \cite{Gunton,Bray,Langer}. 
Our results show that a
slender bar represents
an instructive alternative.  First, the 
slender bar dynamics exhibit
locality of mass transport. Second, two different
dynamical length scales appear. Neither of 
these two properties is present in the single droplet
dynamics.

In conclusion of this section we 
cite earlier papers \onlinecite{Turkevich+Family} where analytical
solutions for the Laplacian {\it growth}
problem were obtained in a slender bar geometry.
Differences in motivation (diffusion-limited
growth versus diffusion-limited coarsening)
lead to important differences between these models and our model.
First, the boundary condition far from the bar in Refs. 
\onlinecite{Turkevich+Family} corresponded to a non-zero 
flux of material, while this flux is zero in our model. 
Second, papers \cite{Turkevich+Family}
did not account for
surface tension at the bar interface.

\section{``Fractal Skeleton" Scenario of Coarsening}
\label{SM}

As it is evident from Fig.~(\ref{fig0}), a two-dimensional
DLA cluster preserves,
in the process of coarsening, its branching structure.
We will characterize
the typical cluster branch at time $t$ by a
width $a(t)$ and length
$b(t)$. Later on we will identify $a(t)$ and $b(t)$ with
the two time-dependent
correlation distances $l_1(t)$ and $l_2(t)$,
respectively. We are interested in
the coarsening regime when
the total mass (area) of the cluster is almost constant.
Define the skeleton of the cluster (at fixed time) by tending all
branches widths to zero. Coarsening of a FC
in our scenario
involves
disappearance of the shortest branches of the cluster
and
an increase of the width of the remaining branches so
that the cluster mass remains constant.
We assume that the cluster {\it skeleton},
rather than the cluster itself,
preserves its fractal structure,
with the same fractal dimension as at $t=0$,
in the process of coarsening \cite{box}. The
lower cutoff of the fractal skeleton is the typical
branch {\it length}, and it grows with time. The upper
cutoff remains constant. Another important assumption is that
each individual daughter branch evolves like a single slender bar
of the previous Section, until the time when
its aspect ratio becomes of order unity.
Then this branch rapidly shrinks and disappears,
``injecting" its material into the parental
branch.

Now Eq.~(\ref{B5}) can be interpreted in the following way.
Because of its shortening, the branch with the width $a_0$ and length $b_0$
has a lifetime of order
\begin{equation}
t_{0}\sim a_0 b_0^2\ .
\label{M1}
\end{equation}
Obviously, only those branches which lifetime is
larger than $t$ will survive by time $t$. Therefore, at time $t$
the typical remaining branches satisfy the scaling relation
\begin{equation}
a(t) \,b^2 (t) \sim t\ .
\label{M1a}
\end{equation}
The same scaling relation apparently holds for
the two
correlation lengths $l_1(t)$
and $l_2(t)$ [see  Eq. (\ref{LB3})]. This
indicates
that $a(t)$ and $b(t)$
can indeed be identified with  $l_1(t)$ and $l_2(t)$, respectively.

The total mass of the cluster can be
estimated as $ab\, (L/b)^D$, and
this quantity must be constant and equal to the initial mass $M$.
Combining it with
Eq.~(\ref{M1a}), we arrive at the following scaling
relations:
\begin{eqnarray}
 a(t) \sim l_1(t)
\sim \left(\frac{M}{L^D}\right)^{\frac{2}{D+1}}\,t^{\frac{D-1}{D+1}}\,,
\label{M3}\\
b(t) \sim l_2(t)
\sim \left(\frac{L^D}{M}\right)^{\frac{1}{D+1}}\, t^{\frac{1}{D+1}}\,.
\label{M4}
\end{eqnarray}
For the cluster perimeter we obtain
\begin{equation}
P (t) \sim b\, \left(\frac{L}{b}\right)^D \sim
M^{-\frac{D-1}{D+1}}\, L^{\frac{2D}{D+1}}\, t^{-\frac{D-1}{D+1}}\ .
\label{M5}
\end{equation}
The different dynamical exponents 
obtained for $a(t)$ and $b(t)$ explain breakdown
of dynamical scale invariance. Notice that
the absolute values of the
exponents for $P(t)$ and $a(t)$
coincide. In the limiting case of $D=2$ {\it all} the exponents
coincide to give the Lifshitz-Slyozov value
$1/3$, so that dynamical
scale invariance is restored. This limiting case describes
a quench through the critical point, where convoluted
percolating interfaces (for which $D=2$)
are observed \cite{RogersDesai,Tomita,Jeppesen}.

In addition, we can give a robust description of
the scaling behavior of the ``solute" mass outside the cluster.
The ``injection" processes outlined above
causes undulations of the interface
of the parent
branches (clearly seen in Fig.~\ref{fig0}). In their turn, diffusion
of the ``solute"  and surface tension tend to erase the
undulations. It is natural to assume
that, by time $t$, undulations with wavelengths
smaller than $\sim t^{1/3}$ have already
disappeared \cite{local}. Hence, at time $t$ the 
typical wavelength (or curvature radius) of the branch undulation is
of order $t^{1/3}$. [Notice that, for
$D<2$,
this curvature radius grows with time slower than $b(t)$.]  The ``solute"
density $\rho$ around the curved
interface is of order
$t^{-1/3}$ due to the Laplacian screening of ``bays" by ``capes".
Solute with this density will be found within
a distance of the order of the
diffusion length $l_d \propto t^{1/2}$ from the cluster interface.
Outside of this region
the density will very small.
The area of this region
can be estimated as
\begin{equation}
S_d\sim l_d^2\, \left(\frac{L}{l_d}\right)^D\propto
t^{\frac{2-D}{2}} \,.
\label{add}
\end{equation}
Multiplying this quantity by the ``solute" density $t^{-1/3}$, we
obtain a dynamical
scaling relation for the ``solute" mass outside the
cluster:
\begin{equation}
M_s \propto t^{\frac{4-3D}{6}}\,.
\label{M6}
\end{equation}
Notice that for $D=2$ the Lifshitz-Slyozov
scaling $t^{-1/3}$ {\it for the
solute mass} \cite{LS} is recovered. A striking prediction of Eq. (\ref{M6})
is the change of sign of the dynamical exponent at $D=4/3$. 
For $D<4/3$
the ``solute" mass should continue {\it increasing} 
with time until the time when the dissolving fractal degrades and 
the ``fractal skeleton" model becomes
inapplicable. It should
be noted that estimate (\ref{M6}) is quite robust, as 
it is independent of most of the assumptions of the ``fractal skeleton"
scenario. For example, it does not use the values of
the first two dynamical exponents and only assumes
that they are less than $1/2$ (that is, 
the length and width of the branches grow in time
slower than the diffusion length $l_d$).

Now we are in a position to compare the predictions
of the
``fractal skeleton" scenario with our simulations of DLA coarsening.
This comparison is made in
Table 1 for $D=1.70$. The agreement is quite reasonable,
in view of the
uncertainty range of the exponents found numerically and
simplicity of the scenario.

%%%%%%%%%%%%table%%%%%%%%%%%%%%%%%%%%%
\begin{table}
\caption{Dynamical exponents: scenario vs. simulation
\label{TT}}
\begin{tabular}{c|ccc}
        & exponent      & exponent    & exponent   \\
Quantity&from scenario & from corrected  & from pure \\
        &               & power-law fit
                                      & power-law fit    \\
\hline
$l_1$  &    0.26 &  0.26  & 0.20 \\
$l_2$  &    0.37 &  ---   & 0.39 \\
$P$    &   -0.26 &  -0.26 & -0.20 \\
$M_s$  &   -0.18 &  -0.24 & -0.16 \\
\end{tabular}
\end{table}
\nopagebreak
\section{Summary and Discussion}

We investigated the bulk-diffusion-controlled
coarsening of
DLA fractal
clusters as described by the CH-equation. We observed that
long-ranged correlations, introduced by the FC
at $t=0$, define a new intermediate asymptotics (in the terminology of
Ref. \cite{Barenblatt}) 
in the coarsening dynamics: ``fractal coarsening".
In order to reach this asymptotics, one should wait long enough so that
quasi-equilibrium domain walls
are already formed and mass transfer between the cluster
and the majority phase is already weak (the cluster mass
is approximately constant).
On the other hand,
this stage is limited at large times by finite-size
effects [a finite value of $l_2 (t)/L$]. This
intermediate asymptotic stage is quite long (our coarsening
scenario predicts that its duration is of order $L^{D+1}$), and it 
certainly deserves
attention. 

The main result of this work (and of the preceding Letter \cite{CMS})
is breakdown
of dynamical scale invariance, that is
the presence of several dynamical length scales, during the
bulk-diffusion-limited
fractal coarsening. We identified two dynamical length scales
from the evolution of the equal-time
pair correlation function, and a third dynamical length scale
from the evolution of the ``solute" mass outside the cluster.
Breakdown of dynamical scale invariance
is caused by the effective locality of  mass transfer
which manifests itself in the {\it simultaneous} 
(approximate) conservation of
the cluster mass and gyration radius in the process 
of coarsening. Scale-invariant mass-preserving
coarsening would obviously 
require shrinking of the FC \cite{Sempere}, while no shrinking
is observed in this system.

Looking for a simple scenario of coarsening with a broken
dynamical scale invariance, we investigated an auxiliary problem of
the dynamics of a single dumbbell-shaped domain
and found ``unusual" dynamical exponents 
$1/2$ and $1/4$. 
Locality of mass transfer is present already in the single-dumbbell
dynamics.
We suggested
a simple scenario of fractal coarsening in diffusion-controlled systems
with a conserved
order parameter. It
postulates 
a fractal skeleton with an invariable fractal dimension and
employs the dumbbell model for the dynamics of individual branches. 
In addition, a robust estimate of the ``solute mass" exponent
is obtained, and a qualitative change in the ``solute mass" dynamics
is predicted at $D=4/3$. Theoretical predictions
are in reasonable agreement with numerical
simulations. 

Much more work is needed, however, before a more complete understanding
of the fractal coarsening emerges. A moderate
fractal range
of the DLA realizations that we worked with
made it difficult to
obtain sharp estimates of the dynamical exponents.
More extensive numerical simulations
would increase the scaling range and 
test the values
of the dynamical exponents.

How can one put the ``fractal skeleton" scenario of coarsening under
additional tests? The scenario
gives very definite
predictions of the dynamical exponents
in terms of the (invariable) fractal dimension of the cluster
skeleton. In general, fractal dimension is only one
characteristics of a geometrical set. Therefore,  
one direct test would involve 
simulations of coarsening of a {\it different} random
FC with branching structure that
has the same 
fractal dimension as DLA. Further tests would involve
simulations with random
FCs having branching structures with
{\it different} (tunable) 
fractal dimensions. Simulations of this type were performed
in Ref. \cite{Jullien} for the case of edge-diffusion-controlled
fractal coarsening. In our case of bulk-diffusion-controlled coarsening
these tests will check, in particular, the prediction of a qualitative change
in the ``solute mass" dynamics at $D_{crit}=4/3$.

Finally, a comparative
investigation of different {\it mechanisms} of fractal coarsening
(edge diffusion 
versus bulk diffusion, local conservation
versus global conservation, etc.) is
needed if we want to address the question about possible
universality classes of fractal
coarsening.

\acknowledgments

We are very grateful to Avner Peleg and Azi Lipshtat for help, and to
Roman Kris for useful discussions.
This work was supported in part
by a grant from Israel Science Foundation, administered
by the Israel Academy of Sciences and Humanities, and by the
Russian Foundation
for Basic Research (grant No. 99-01-00123).


\begin{references}
\bibitem{Gunton} J.D. Gunton, M. San Miguel, and P.S. Sahni, in {\it Phase
Transitions and Critical Phenomena}, edited by C. Domb and J.L. Lebowitz
(Academic Press, New York, 1983), Vol. 8, p. 267.
\bibitem{Bray} A.J. Bray, Adv. Phys. {\bf 43}, 357 (1994).
\bibitem{Langer} J.S. Langer, Rev. Mod. Phys. {\bf 52}, 1 (1980); and in
{\it Chance and Matter}, edited by J. Souletie, J. Vannimenus, and
R. Stora (Elsevier, Amsterdam, 1987).
\bibitem{Kessler} D.A. Kessler, J. Koplik, and H. Levine, Adv.
Phys. {\bf 37}, 255 (1988).
\bibitem{Brener} E.A. Brener and V.I. Mel'nikov, Adv. Phys. {\bf 40},
53 (1991).
\bibitem{Mineev} M.B. Mineev-Weinstein, in {\it Fluctuations and Order.
The New Synthesis} (Springer, New York, 1996), p. 239.
\bibitem{BMT} E. Brener, H. M\"{u}ller-Krumbhaar, and D. Temkin, Phys. Rev.
E {\bf 54}, 2714 (1996).
\bibitem{Irisawa95} T. Irisawa, M. Uwaha and Y. Saito, Europhys. Lett.
{\bf 30}, 139 (1995).
\bibitem{CMS} M. Conti, B. Meerson and P. Sasorov,
Phys. Rev. Lett. {\bf 80}, 4693 (1998).
\bibitem{fingering} K.V. McCloud and J.V. Maher, Physics Reports
{\bf 260}, 139 (1995).
\bibitem{Sempere} R. Semp\'{e}r\'{e}, D. Bourret, T. Woignier, J. Phalippou
and R. Jullien, Phys. Rev. Lett. {\bf 71}, 3307 (1993).
\bibitem{Hinic} I. Hinic, Phys. Stat. Sol. (a) {\bf 144}, K59, 1238 (1994).
\bibitem{Jullien} N. Olivi-Tran, R. Thouy and R. Jullien, J. Phys.
I {\bf 6}, 557 (1996).
\bibitem{Streitenberger} P. Streitenberger, D. F\"{o}rster and
P. Veit, Fractals {\bf 5}, Suppl. Issue, 5 (1997).
\bibitem{Crooks} G.E. Crooks, B. Ostrovsky and Y. Bar-Yam, Phys. Rev. E
{\bf 60}, 4559 (1999).
\bibitem{Witten} T.A. Witten, Jr. and L.M. Sander, Phys. Rev. Lett. {\bf 47},
1400 (1981).
\bibitem{LS} I.M. Lifshitz and V.V. Slyozov, J. Phys. Chem. Solids
{\bf 19}, 35 (1961).
\bibitem{RogersDesai} T.M. Rogers and R.C. Desai, Phys. Rev. B {\bf 39}, 11956
(1989).
\bibitem{Tomita} H. Tomita, Prog. Theor. Phys. 85, {\bf 47} (1991).
\bibitem{Jeppesen} C. Jeppesen and O.G. Mouritsen, Phys. Rev. B {\bf 47},
14724 (1993).
\bibitem{Huse} D. A. Huse, Phys. Rev. B, {\bf 34}, 7845 (1985).
\bibitem{Vicsek} T. Vicsek, {\it Fractal Growth Phenomena} (World Scientific,
Singapore, 1992).
\bibitem{Toyoki} H. Toyoki and K. Honda, Phys. Lett. {\bf 111}, 367 (1985).
\bibitem{remark} The arrest of dissolution requires also that
the fractal dimension of the cluster be larger than 4/3 
[see Eq. (\ref{M6}) and the subsequent discussion].
\bibitem{Pego} R.L. Pego, Proc. R. Soc. Lond. A {\bf 422}, 261 (1989).
%\bibitem{morecomplete} A more general sharp-interface model \cite{CMS}
%does not assume that the relevant length scales are smaller
%than the diffusion length $l_d \sim t^{1/2}$, but it is much more difficult to
%handle analytically.
\bibitem{Turkevich+Family} L.A. Turkevich and H. Scher, Phys. 
Rev. Lett. {\bf 55}, 1026 (1985); F. Family and H.G. Hentschel, Faraday Disc.
Chem. Soc. {\bf 83}, 139 (1987).
\bibitem{Jackson} J.D. Jackson, {\it Classical Electrodynamics}
(Wiley, New York, 1975), p. 77.
\bibitem{box} Speaking about the fractal dimension of the skeleton,
we mean its box-counting dimension \cite{Vicsek} and assume
that it
is equal to the mass fractal dimension defined by
the pair correlation function.
\bibitem{local} By assuming the  Lifshitz-Slyozov 
value of the corresponding
dynamical exponent, we treat the undulation dynamics as almost local
(that is, independent of the global structure of the cluster and, 
in particular,
of the fractal dimension of its skeleton).
\bibitem{Barenblatt} G.I. Barenblatt, {\it Scaling, Self-Similarity, and
Intermediate Asymptotics} (Cambridge Univ. Press, Cambridge, 1996).

\end{references}
\end{document}